\documentclass{article}
\usepackage{dcase2019,amsmath,graphicx,url,times,booktabs, tabularx, multirow}
\usepackage{caption}
\usepackage{subcaption}

\title{DD-CNN: Depthwise Disout Convolutional Neural Network for Low-complexity Acoustic Scene Classification}

\name{Jingqiao Zhao$^{1}$,
       Zhen-Hua Feng$^{2,3}$,
       Qiuqiang Kong$^{4}$, 
       Xiaoning Song$^{1}$,
       Xiao-Jun Wu$^{1}$, 
       }
 \address{$^1$ School of Artificial Intelligence and Computer Science, Jiangnan University, Wuxi, China\\       
         $^2$ Department of Computer Science, University of Surrey, Guildford, UK \\
$^3$Centre for Vision, Speech and Signal Processing, University of Surrey, Guildford, UK\\
         $^4$ ByteDance AI Lab, Shanghai, China\\
         zhao\_jing\_qiao@163.com, z.feng@surrey.ac.uk, kongqiuqiang@bytedance.com\\ x.song@jiangnan.edu.cn, wu\_xiaojun@jiangnan.edu.cn \\
  }

\begin{document}

\ninept
\maketitle

\begin{sloppy}

\begin{abstract}
This paper presents a Depthwise Disout Convolutional Neural Network (DD-CNN) for the detection and classification of urban acoustic scenes. Specifically, we use log-mel as feature representations of acoustic signals for the inputs of our network. In the proposed DD-CNN, depthwise separable convolution is used to reduce the network complexity. Besides, SpecAugment and Disout are used for further performance boosting. Experimental results demonstrate that our DD-CNN can learn discriminative acoustic characteristics from audio fragments and effectively reduce the network complexity. Our DD-CNN was used for the low-complexity acoustic scene classification task of the DCASE2020 Challenge, which achieves 92.04\% accuracy on the validation set.
\end{abstract}


\section{Introduction}
\label{sec:intro}
Environmental sounds carry rich information that can be used for a wide spectrum of intelligent applications, such as Internet of Things services~\cite{hossain2015towards} and surveillance systems~\cite{dufaux2008scrambling}. Acoustic Scene Classification (ASC) \cite{barchiesi2015acoustic, mesaros2018multi} is a branch of computational auditory scene analysis that aims to detect and classify surrounding sounds. To be more specific, the aim of ASC is to classify a test recording into one of the predefined classes that characterizes the environment in which it was recorded.

The existing ASC algorithms can be roughly divided into two main categories: classical methods and deep learning based methods. In earlier studies, a variety of speech recognition methods were successfully applied to ASC using features such as Mel frequency cepstral coefficients and normalized spectrum features~\cite{Aucouturier2007TheBA}. Well-known classical ASC methods include hidden Markov models (HMMs)~\cite{jakovljevic2017hidden}, Support Vector Machine (SVM)~\cite{Geiger2013LargescaleAF} and decision trees~\cite{Phan2016LabelTE}. In recent years, deep learning has become the mainstream for ASC. Similar to speech recognition, deep neural networks usually use time-frequency representations as inputs for ASC. In this category, different network architectures, such as Fully Connected Networks (FCN), Convolutional Neural Networks (CNN) and Recurrent Neural Networks (RNN), have been applied to acoustic scene classification ~\cite{abesser2020review, dorfer2018acoustic, xu2018mixup}. In recent years, CNN have achieve promising results in various ASC benchmarks~\cite{Chen2019IntegratingTD, Wang2019HybridCT}.

To benchmark an ASC method, the DCASE challenges \cite{mesaros2017dcase, mesaros2018multi} might be the most well-known platform, which is organized annually.
The DCASE2020 ASC challenge consists of two sub-tasks: ASC with multiple devices and low-complexity ASC. In this work, we focus on the second sub-task~\footnote{http://dcase.community/challenge2020}. This task consists of three types of sounds (indoor, outdoor and transportation) captured by the same device to form the development and evaluation datasets. A two-layer Convolutional Neural Network (CNN) is provided by the organizers as the baseline approach. It has two convolutional blocks, each consists of Convolution, BatchNorm, Activation, MaxPooling and Dropout layers. Last, a densely connected decision-making layer is used to perform the classification task.

To achieve robust ASC with low computational complexity, we propose a Depthwise Disout CNN (DD-CNN) using log-mel spectrum features. The key innovative elements of the proposed DD-CNN model include: 1) We use the depthwise separable convolution to reduce the network complexity while maintaining the network performance. 2) We apply SpecAugment for training data augmentation and use Disout to address the network overfitting issue. The average accuracy of the baseline method is 87.3\% for the DCASE2020 low-complexity ASC task. Our DD-CNN significantly improves the performance over the baseline method by achieving 92.04\% in terms of the average accuracy. 

\begin{figure}[t]
  \centering
  \centerline{\includegraphics[width=.9\columnwidth]{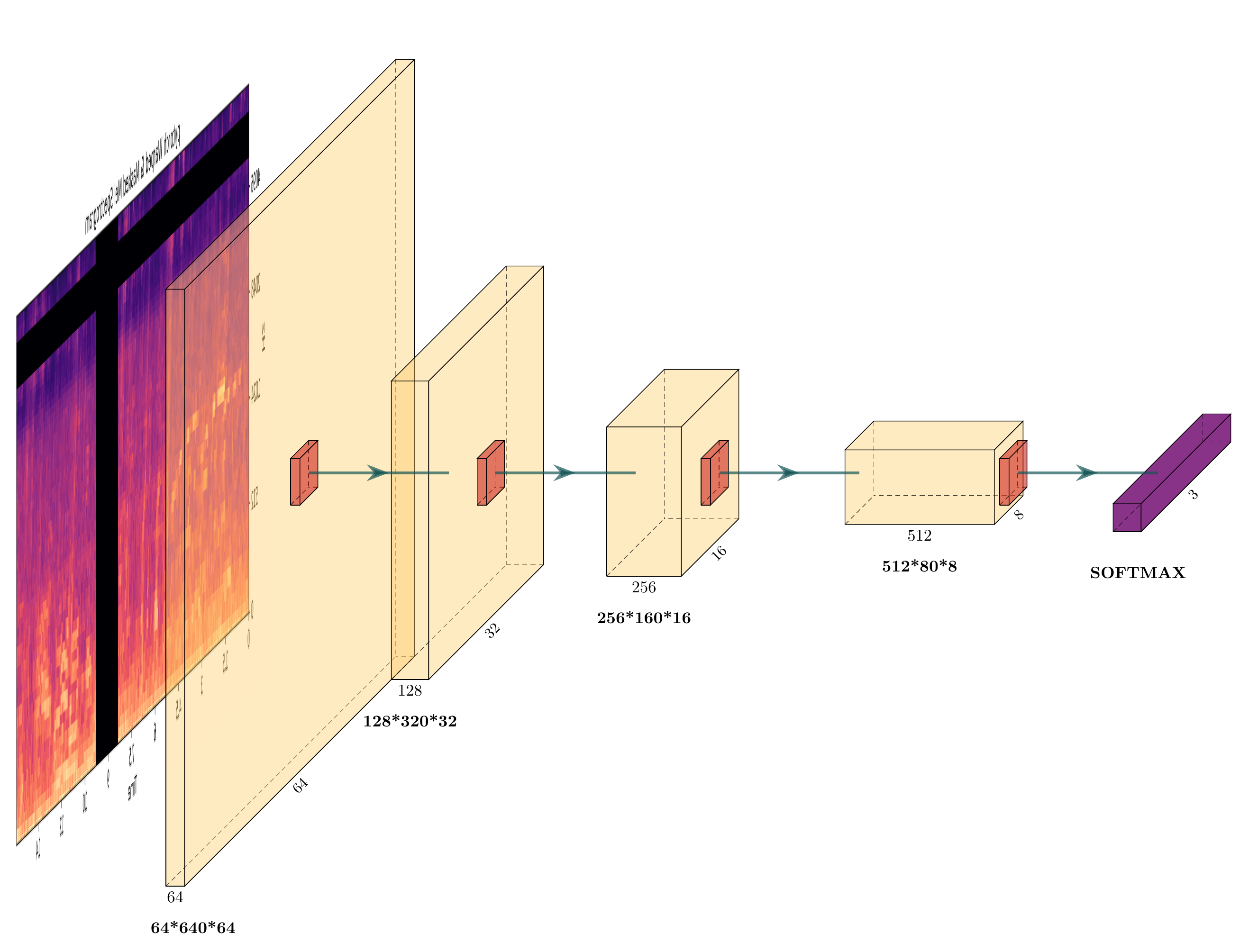}}
  \caption{Network architecture of our five-layer CNN (CNN-5) model with average pooling.}
  \label{fig1}
\end{figure}

\section{The Proposed Method}
\label{sec:method}
\subsection{CNN-based ASC}
CNNs have been widely used in many computer vision and pattern analysis tasks, which achieve state-of-the-art performance in a wide range of image and video analysis tasks, such as detection~\cite{he2017mask, feng2017face}, localization~\cite{feng2019rectified, feng2019mining}, tracking~\cite{xu2020afat} and classification~\cite{krizhevsky2012imagenet,he2016deep}. A traditional CNN architecture usually consists of several convolutional layers, each has multiple filters to process the feature map output from the previous layer. Filters can capture local patterns of an input feature map, such as edges in lower levels and object contours in higher levels. CNNs have also been used for many audio classification and sound event detection tasks~\cite{gemmeke2017audio,kong2018dcase}, using inputs such as log-mel spectrum. However, previous CNN-based ASC methods do not consider much in the computational complexity of a network.

For the low-complexity ASC task, we first use a five-layer CNN (CNN-5) architecture that is similar to AlexNet~\cite{krizhevsky2012imagenet}. The architecture and detailed configuration of our CNN-5 are shown in Figure.~\ref{fig1} and Table~\ref{table1}, respectively. As shown in Figure.~\ref{fig2}, our CNN-5 achieve good performance on the training set. However, the number of parameters of CNN-5 is much more than 500KB (Table~\ref{table1}). 

We reduce the network complexity of CNN-5 by removing two convolutional layers and obtain a simple CNN-3 model which model size is smaller than 500KB. However, as shown in Figure.~\ref{fig2}, the performance of CNN-3 is degraded significantly, in terms of training accuracy. To address this issue, we propose a novel Depthwise Disout CNN (DD-CNN) model that uses depthwise separable convolution to reduce the complexity of a network while maintaining the performance of the network. In addition, we use Disout to further improve the generalization ability of our DD-CNN model. The network architecture and detailed configuration of our DD-CNN model are shown in Figure.~\ref{fig_ddcnn} and Table~\ref{table2}, respectively.
\begin{figure}[t]
  \centering
  \includegraphics[width=\columnwidth]{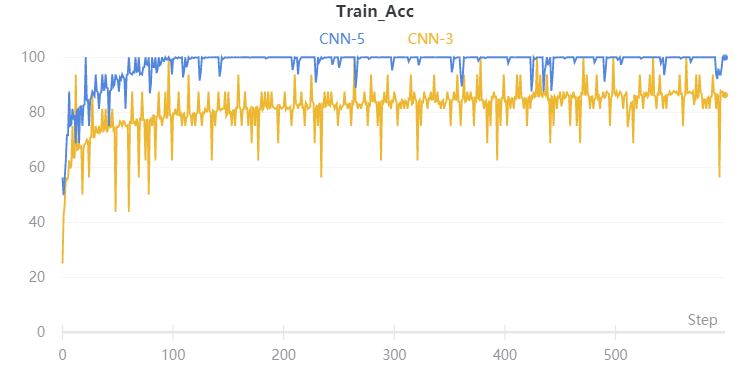}
  \caption{The performance of CNN-5 and CNN-3 on the training set, in terms of accuracy.}
  \label{fig2}
\end{figure}

\begin{figure}[t]
  \centering
  \centerline{\includegraphics[width=\columnwidth]{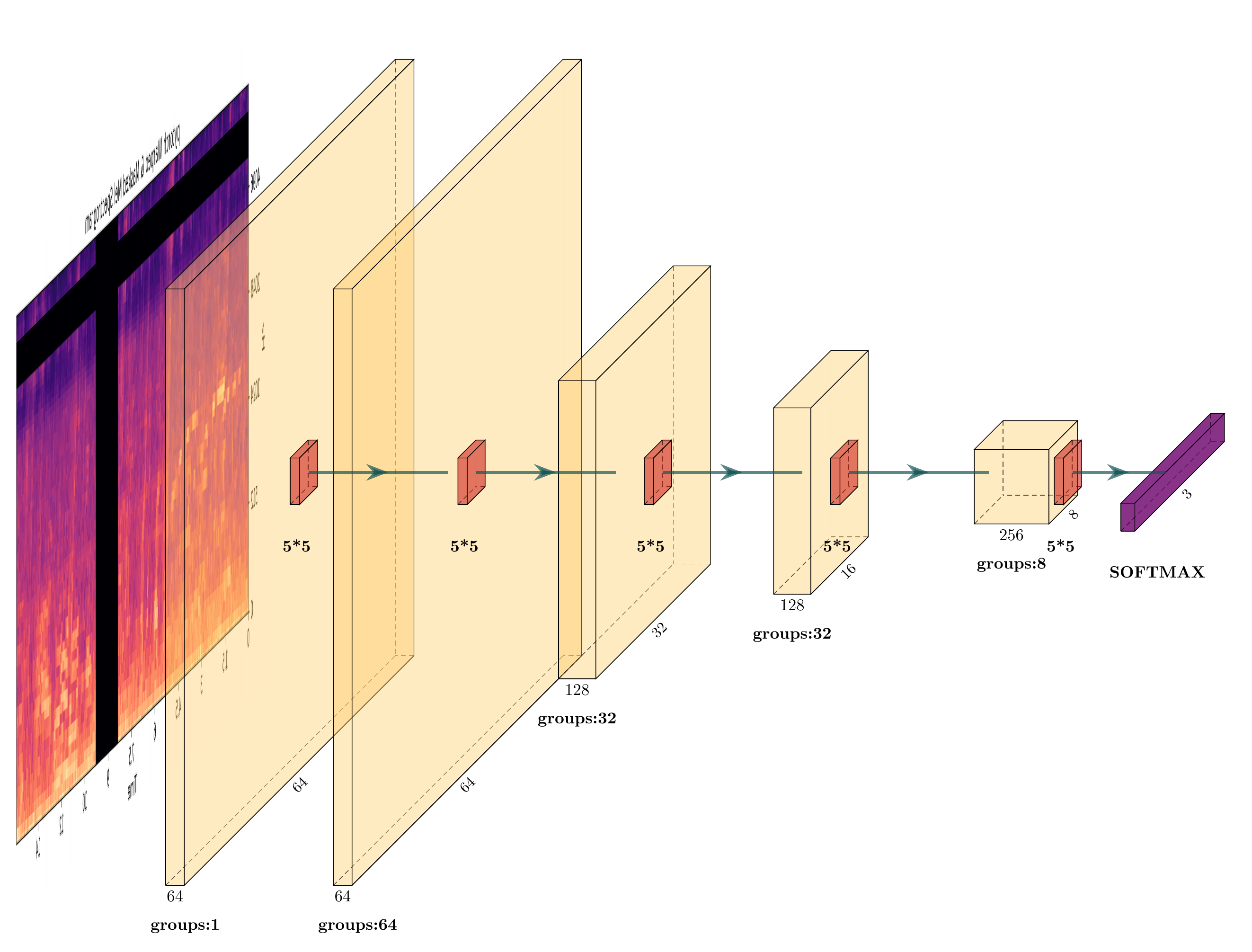}}
  \caption{The network architecture of the proposed DD-CNN method.}
  \label{fig_ddcnn}
\end{figure}

\begin{figure}[!t]
  \centering
  \includegraphics[width=.4\columnwidth]{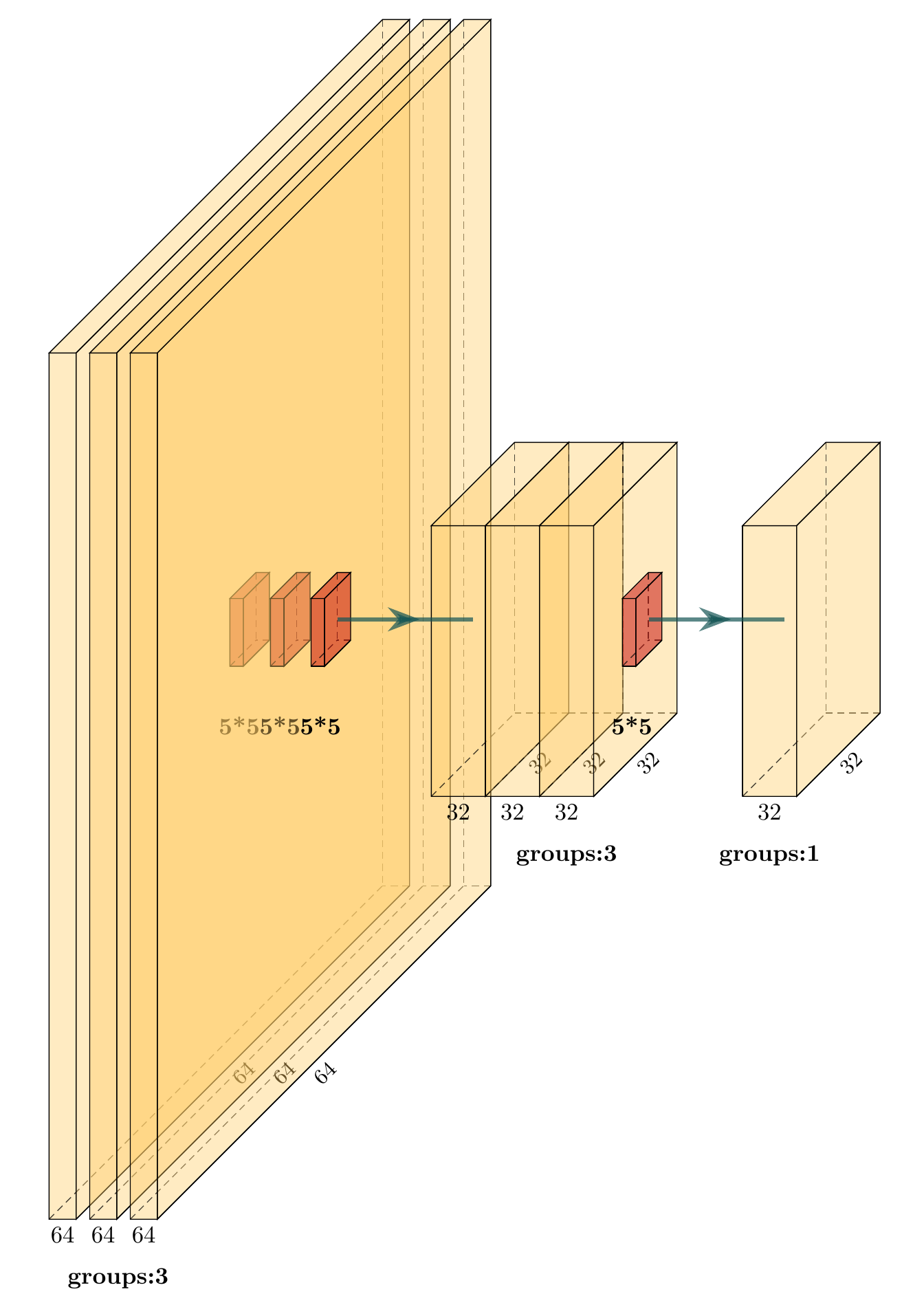}
  \caption{Depthwise separable convolution.}
  \label{fig4}
\end{figure}

\begin{table}[t]
\centering
\caption{Configuration of the CNN-5 architecture. The total number of model parameters is 4,305,859.}
\begin{tabular}{lll}
\hline
\textbf{Layer}  & \textbf{Output Shape} & \textbf{\# Params}  \\
\hline
Conv2d-1      & {[}-1, 64, 640, 64{]}  & 1,600     \\
Conv2d-1      & {[}-1, 64, 640, 64{]}  & 1,600     \\
BatchNorm2d-2 & {[}-1, 64, 640, 64{]}  & 128       \\
Conv2d-3      & {[}-1, 128, 320, 32{]} & 204,800   \\
BatchNorm2d-4 & {[}-1, 128, 320, 32{]} & 256       \\
Conv2d-5      & {[}-1, 256, 160, 16{]} & 819,200   \\
BatchNorm2d-6 & {[}-1, 256, 160, 16{]} & 512       \\
Conv2d-7      & {[}-1, 512, 80, 8{]}   & 3,276,800 \\
BatchNorm2d-8 & {[}-1, 512, 80, 8{]}   & 1,024     \\
Linear-13     & {[}-1, 3{]}            & 1,539    \\
\hline
\end{tabular}
\label{table1}
\end{table}

\begin{table}[!h]
\caption{Configuration of the proposed DD-CNN architecture. The total number of model parameters is 127, 491. }
\centering
\begin{tabular}{lll}
\hline
\textbf{Layer}  & \textbf{Output Shape} & \textbf{\# Params}  \\
\hline
Conv2d-1           & {[}-1, 64, 640, 64{]}   & 1664     \\
BatchNorm2d-2      & {[}-1, 64, 640, 64{]}   & 128      \\
Conv2d-3           & {[}-1, 64,   640, 64{]} & 1664     \\
BatchNorm2d-4      & {[}-1, 64, 640, 64{]}   & 128      \\
Conv2d-5           & {[}-1, 128, 320, 32{]}  & 6528     \\
BatchNorm2d-6      & {[}-1, 128, 320, 32{]}  & 256      \\
Conv2d-7           & {[}-1, 128, 160, 16{]}  & 12928    \\
BatchNorm2d-8      & {[}-1, 128, 160, 16{]}  & 256      \\
Conv2d-9           & {[}-1, 256, 80, 8{]}    & 102656   \\
BatchNorm2d-10     & {[}-1, 256, 80, 8{]}    & 512      \\
Disout-11          & {[}-1, 256, 40, 4{]}    & 0        \\
LinearScheduler-12 & {[}-1, 256, 40, 4{]}    & 0        \\
Linear-13          & {[}-1, 3{]}             & 771    \\
\hline
\end{tabular}
\label{table2}
\end{table}

Depth-level separable convolution is a decomposable convolution operation that can be broken down into two smaller operations: depthwise convolution and pointwise convolution~\cite{howard2017mobilenets}. Depthwise convolution is different from standard convolution. For the conventional convolution, filers are applied to all the input channels. In contrast, depthwise convolution uses different convolution filters for different input channels, as shown in Figure.~\ref{fig4}. The last convolution layer is applied to all the channels, so the depth separable convolution is equal to the standard convolutional operation. Pointwise convolution is a normal convolution, except that it uses 1x1 convolutions. For depthwise separable convolution, it first uses depthwise convolution of different input channels separately, and then uses pointwise convolution to combine all the above outputs. With depthwise separable convolution, we can significantly reduce the parameters and computational complexity of a network while maintaining its performance, as shown in Table~\ref{table2}.

By using depthwise separable convolution, the network architecture of our DD-CNN is shown in Figure.~\ref{fig_ddcnn}. More details of the configuration of each layer of our DD-CNN is shown in Table~\ref{table2}. For each convolutional layer, we set the kernel size to 5, padding size to 2 and stride to 1. The fully-connected layer with Disout is used for output.

\subsection{Disout}
Dropout is an important technique for effective deep learning, which has been widely used in various machine learning frameworks. Dropout brings positive impacts to deep-learning-based computer vision, natural language processing, speech recognition and other tasks. Dropout can improve the performance of a trained network in terms of generalization capability and robustness by addressing the overfitting problem.

\begin{figure}[t]
\centering
\begin{subfigure}[h]{\columnwidth}
\includegraphics[width=\columnwidth]{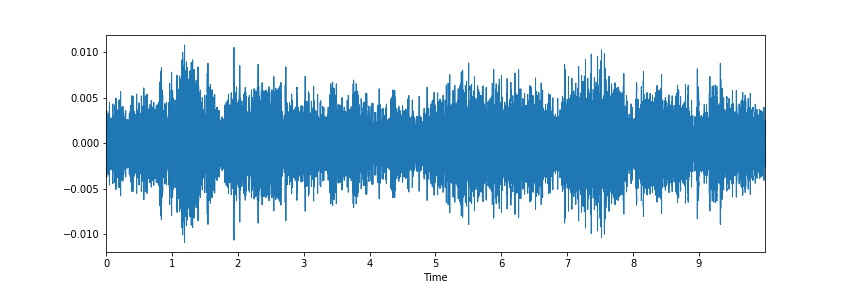}
\caption{Original waveform}
\label{fig5_1}
\end{subfigure}
\\

\begin{subfigure}[h]{\columnwidth}
\includegraphics[width=\columnwidth]{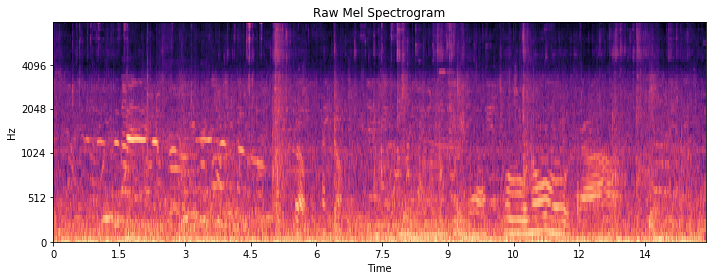}
\caption{The corresonding log-mel spectrum}
\label{fig5_2}
\end{subfigure}

\begin{subfigure}[h]{\columnwidth}
\includegraphics[width=\columnwidth]{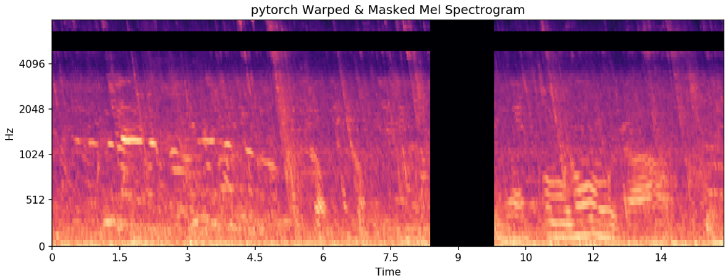}
\caption{SpecAugment}
\label{fig5_3}
\end{subfigure}

\caption{SpecAugment for log-mel spectrum map}
\label{fig5}
\end{figure}

Based on dropout, Tang \textit{et. al.} proposed a new technique, namely Disout~\cite{tang2020beyond}. This method enhances the generalization ability of deep neural networks by studying feature plot disturbances. Based on the Rademacher complexity of the middle layer of a network, the generalization error upper bound of a given deep neural network is determined. The introduction of perturbs into feature maps in Disout can reduce the complexity of the network's Rademacher, thereby improving its generalization capability. Disout not only outperforms the classical dropout method in traditional visual tasks but also achieves promising results in natural language processing and voice processing tasks. In our DD-CNN, we use Disout for the last fully-connected layer to improve the generalization capacity of the network.

\subsection{SpecAugment}
Without sufficient training data, it is crucial to apply data augmentation to the existing training samples thus to improve the performance of a learning-based method by better exploiting the data in our hand. In sound recognition, traditional data augmentation methods include deformation of sound waves and background noise jetting. Different data augmentation methods are applied to each individual training sample. With data augmentation, we can train a network with better performance by synthesizing new training samples from the original ones. However, existing data augmentation methods, such as ASRs~\cite{moon2015rnndrop}, increase computational complexity and often require additional data.

To train a CNN model for ASC, we first transform sound waves to a spectrum map as our training data. Data augmentation methods are typically applied to sound waves and then transformed into spectrum maps. For each sample, newly generated sound waves have to be converted to spectrum maps, increasing the amount of calculation. In our approach, we directly perform data augmentation on spectrum maps, rather than the original sound data. Because our data augmentation method is applied directly to input spectrum maps, it can be added dynamically in real time without spending more time on converting sound waves to spectrum maps, which slow down the network training significantly.

To be more specific, we use SpecAugment~\cite{park2019specaugment} to modify spectrum maps by distorting time domain signal, masking the frequency domain channel and the time domain channel. This data augmentation method can be used to increase the robustness of the trained network to combat deformations on the time domain and partial fragment loss on the frequency domain. In Figure.~\ref{fig5}, we give an example of SpecAugment. 

\section{Experimental Results}
\subsection{The DCASE2020 ASC Dataset}
The DCASE2020 dataset consists of 10 classes sounds captured in airport, shopping mall, metro station, pedestrian street, public square, street traffic, tram, bus, metro and park. This challenge provides two datasets, development and evaluation, for algorithm development. The sub-task B of TAU Urban Acoustic Scenes 2020 dataset contains 40 hour audio recordings which are balanced between classes and recorded at 48kHz sampling rate with 24-bit resolution in stereo. Each sound recording was spitted into 10-second audio samples.

\begin{figure}[t]
  \centering
  \includegraphics[width=\columnwidth]{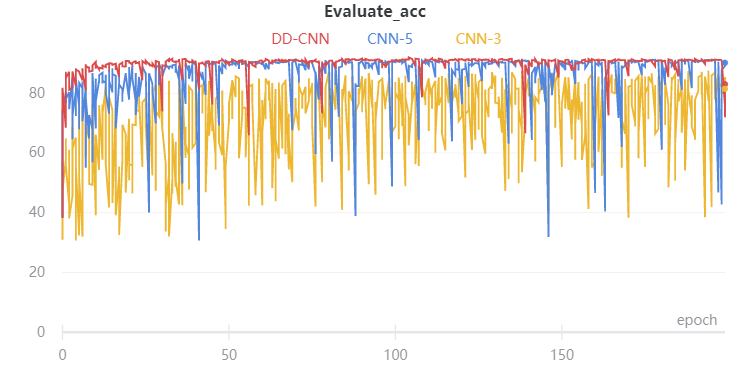}\\
  \caption{The performance of CNN-3, CNN-5 and DD-CNN in terms of accuracy, obtained on the evaluation set.}
  \label{fig6}
\end{figure}

\subsection{Results}
We implemented our CNN-3, CNN-5 and DD-CNN methods using PyTorch. The AdamW optimizer was used for network training, with 200 epochs on an Nvidia RTX 2080ti card. The initial learning rate was set to 0.001 and the batch size was set to 16.

The evaluation performance of three different networks, CNN-3, CNN-5 and DD-CNN, obtained on the evaluation set is shown in Figure.~\ref{fig6}. We can see that the proposed DD-CNN method outperforms CNN-5 significantly in terms of accuracy. However, our DD-CNN has much smaller model parameter size than that of CNN-5. In contrast, by simply reducing the model size of CNN-5 to create a smaller network CNN-3, the performance of CNN-3 is significantly degraded.

\begin{figure}[t]
  \centering
  \includegraphics[width=\columnwidth]{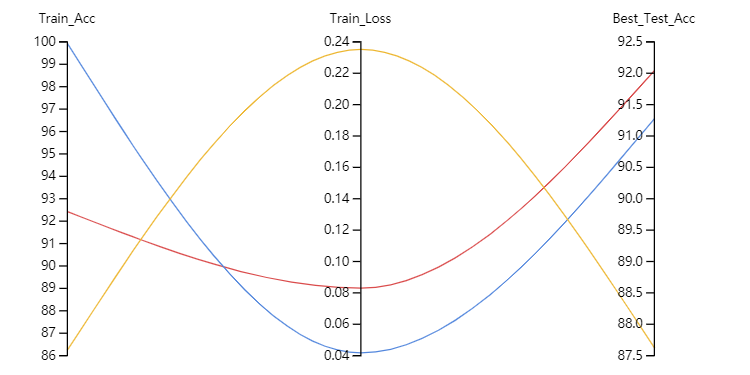}\\
  \caption{The performance of three different networks. Red: our Depthwise Disout CNN, Blue: CNN-5, Yellow: CNN-3.
}
  \label{fig7}
\end{figure}

The final results of the proposed DD-CNN method are reported in Table~\ref{table3}. We can see that our DD-CNN outperforms the baseline approach provided by the organizers for all the categories, in terms of accuracy. Our DD-CNN does not lose accuracy by significantly reducing the model parameters. More importantly, our DD-CNN has better generalization performance than traditional CNN networks as shown in Figure.~\ref{fig7}. We also report the confusion matrix in Figure.~\ref{fig8}.

\begin{table}[t]
\caption{Experimental results under three difference scenarios: indoor, outdoor and transportation.}
\centering
\begin{tabular}{lllll}
\hline
\multirow{2}{*}{\textbf{Scene}} & \multicolumn{2}{c}{\textbf{Baseline}}     & \multicolumn{2}{c}{\textbf{DD-CNN}}       \\
\cline{2-5}
  & \multicolumn{1}{c}{Accuracy} & \multicolumn{1}{c}{Loss} & \multicolumn{1}{c}{Accuracy} & \multicolumn{1}{c}{Loss} \\
\hline
Indoor         & 78.7\%     & 0.863     & 89.1\%     & 0.402    \\
Outdoor        & 88.5\%    & 0.400     & 90.2\%   & 0.296   \\
Transportation & 93.6\%        & 0.236      & 97.3\%    & 0.075    \\
\hline
Average   &    87.0\% &    0.493 &     92.0 \% & 0.257
\\
\hline
\end{tabular}
\label{table3}
\end{table}

\begin{figure}[t]
  \centering
  \includegraphics[width=\columnwidth]{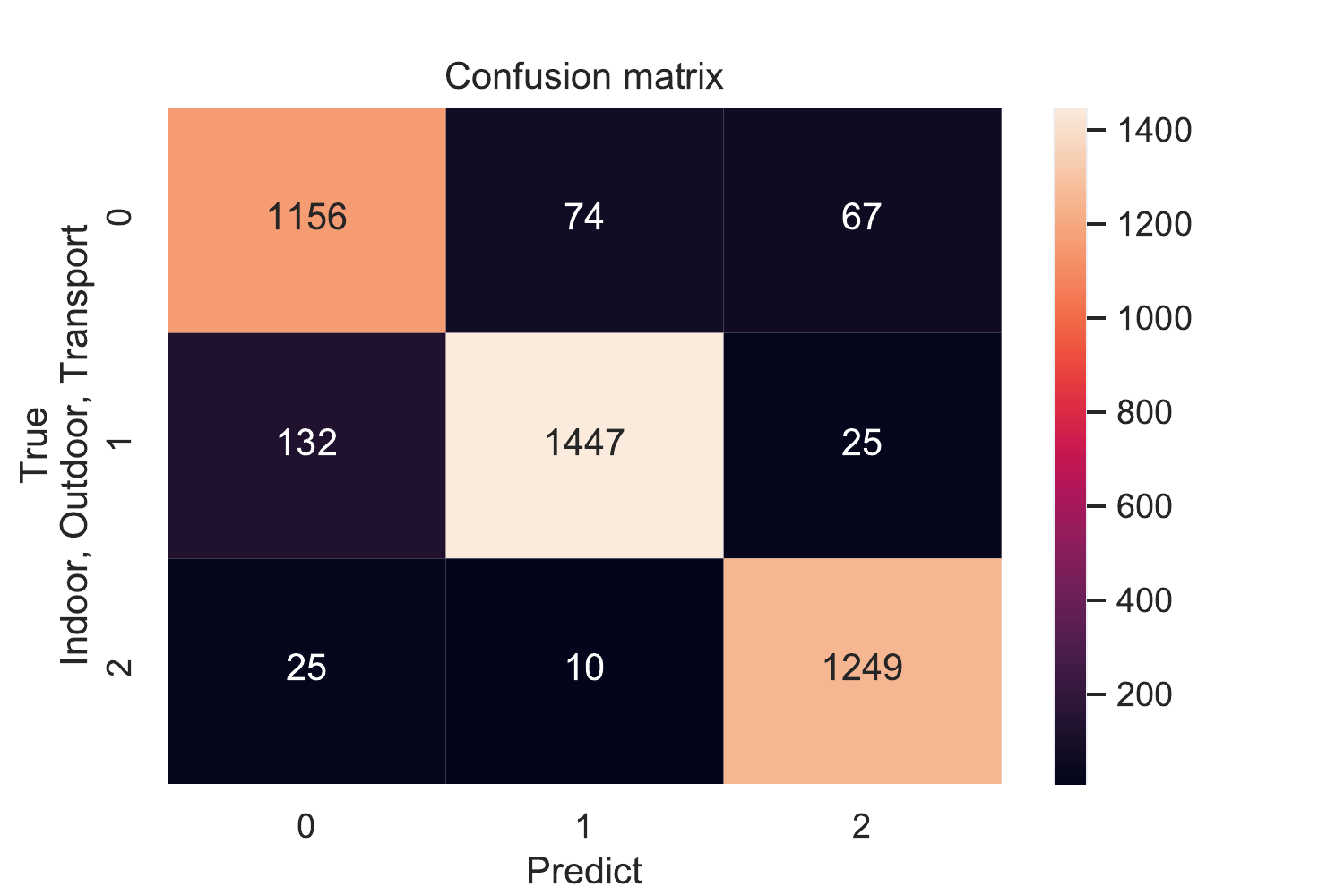}\\
  \caption{The confusion matrix of the best model obtained on the development dataset.}
  \label{fig8}
\end{figure}

\section{Conclusion}
\label{sec:typestyle}
This paper proposed a novel Depthwise Disout CNN (DD-CNN) model for low-complexity acoustic scene classification. In addition, we studied the performance of two traditional CNN models, CNN-3 and CNN-5, with a comparison with our DD-CNN on the low-complexity acoustic scene classification task of the DCASE 2020 challenge. Through our experiments, we find that the deep separable convolution can effectively reduce network parameters while maintaining the performance of a network. In addition, the use of Disout and SpecAugment further improves the generalization capability of our DD-CNN and brings additional performance boosting. In future, we aim to explore cross-mission structures with deep separable convolution networks for CNN-based acoustic scene classification.

\section{ACKNOWLEDGMENT}
This work was supported in part by the National Key Research and Development Program of China (Grant No. 2017YFC1601800), the National Natural Science Foundation of China (61876072, 61902153), the China Postdoctoral Science Foundation (2018T110441) and the Six Talent Peaks Project of Jiangsu Province (XYDXX-012).

\bibliographystyle{IEEEtran}
\bibliography{refs}
%
%
%
%
%
%
%
%
%

\end{sloppy}
\end{document}